\documentclass[%
 reprint,
 amsmath,amssymb,
 aps,
]{revtex4-2}

\usepackage{graphicx}
\usepackage{dcolumn}
\usepackage{bm}
\usepackage{xcolor}
\begin{document}

\preprint{APS/123-QED}

\title{Dark Matter-Baryon Separability Predicts the Dynamics of an Almost-Dark Galaxy}

\author{Oem Trivedi $^{1}$, Abraham Loeb $^{2}$}
\affiliation{$^{1}$Department of Physics and Astronomy, Vanderbilt University, Nashville, TN, 37235, USA}
\affiliation{$^{2}$Astronomy Department, Harvard University, 60 Garden St., Cambridge, 02138, MA, USA}
\email{Email: oem.trivedi@vanderbilt.edu \\ Email : aloeb.cfa@harvard.edu }

\date{\today}

\begin{abstract}
We extend the Dark Matter-Baryon Separability Condition to show that the same framework developed for dark matter deficient galaxies naturally admits a conjugate branch describing preferential baryonic depletion. Using the recently discovered almost-dark galaxy TTT J1237327+143535 as a worked example, we derive a family of dynamical consistency relations parameterized by the unknown progenitor ratio $\mu_i$, including thresholds for $\sigma_{\rm los}$, $M_{\rm dyn}$, the enclosed baryonic fraction and the dynamical mass to light ratio. We further connect the separability framework to the expected globular cluster population, providing an independent consistency test of the inferred halo mass. We also provide some bounds on the baryon ratio using the tidal properties of Virgo Cluster. Future measurements of stellar kinematics, gas content and globular clusters can therefore determine whether this galaxy occupies the positive separability branch and also test whether dark matter deficient and baryon depleted systems can be described within a common framework.
\end{abstract}

\maketitle

The recent discovery of the extremely faint galaxy TTT J1237327+143535 \cite{tttle2026serendipitous} in the Virgo Cluster provides a particularly interesting addition to the population of unusual low surface brightness systems. The object has a central surface brightness of approximately $27.9\,\mathrm{mag\,arcsec^{-2}}$ in the $g'$ band, an effective radius $R_e=0.93\pm0.04\,\mathrm{kpc}$ and a total stellar mass $M_\star=(2.16\pm0.38)\times10^6M_\odot$. Its stellar surface mass density is below $1\,M_\odot\,\mathrm{pc^{-2}}$ even near its centre, while no HI counterpart has been detected in ALFALFA and no H$\alpha$ emission has been reported. This galaxy seems to represent one of the most extreme known examples of an extended galaxy containing remarkably little observable baryonic matter. \\

The designation "almost-dark" is based primarily on the exceptionally low optical surface brightness of the system and does not imply that its dark matter content \cite{dm1Cirelli:2024ssz,dm2Arbey:2021gdg,dm3Balazs:2024uyj,dm4Eberhardt:2025caq,dm5Bozorgnia:2024pwk,dm6Misiaszek:2023sxe,dm7OHare:2024nmr,dm8Adhikari:2022sbh,dm9Miller:2025yyx,dm10Trivedi:2025vry,dm11rubin1970rotation,hdmTrivedi:2025sbe,pbh1pbhzel1966hypothesis,pbh2hawking1971gravitationally,pbh3carr1974black} has been dynamically measured. In particular, there is presently no reported line of sight stellar velocity dispersion for TTT J1237327+143535 and so, no direct determination of its enclosed dynamical mass or dark matter to baryon ratio. Moreover, its association with the Virgo Cluster is physically well motivated but presently relies on an assumed Virgo distance rather than a direct distance measurement. The discovery hence establishes an extremely diffuse and low stellar mass galaxy, but it does not yet establish that the object occupies an unusually dark matter dominated halo. \\

We recently proposed the Dark Matter-Baryon Separability Condition \cite{trilobdm} as a way of translating the existence of dark matter deficient galaxies \cite{df21van2018galaxy,df22danieli2019still,df23wasserman2018deficit,df41van2019second,df42montes2020galaxy,df43li2024rotation,df91keim2026third,df92gannon2023keck,df93keim2025kinematic,df94van2022trail,df95tang2026new,fcc1romanowsky2024candidate,fcc2tang2025unexplained,fcc3buzzo2025new,fcc4buzzo2026dark,Lin:2026rfm} into constraints on the relative incorporation of dark matter and baryons during galaxy formation. Consider an initial system with dark matter mass $M_\chi^i$ and baryonic mass $M_b^i$, and define its initial dark matter to baryon ratio as
\begin{equation}
\mu_i=\frac{M_\chi^i}{M_b^i}
\end{equation}
If a particular formation channel retains or incorporates fractions $\epsilon_\chi$ and $\epsilon_b$ of the initial dark matter and baryonic components, respectively, then the corresponding final masses are
\begin{equation}
M_\chi^f=\epsilon_\chi M_\chi^i
\end{equation}
\begin{equation}
M_b^f=\epsilon_b M_b^i
\end{equation}
The final dark matter to baryon ratio is hence given as
\begin{equation}
\mu_f=\mu_i\frac{\epsilon_\chi}{\epsilon_b}
\end{equation}
This relation is the basic statement of our separability framework. The final composition of the system is controlled not by the absolute amount of either component that is lost, but by their relative retention or incorporation efficiencies. If both components are affected equally, the original mass ratio is preserved, whereas any difference between $\epsilon_\chi$ and $\epsilon_b$ changes the final composition. For a dark matter deficient system one requires $\mu_f\leq\mu_{\rm crit}$, which is equivalent to
\begin{equation}
\frac{\epsilon_\chi}{\epsilon_b}\leq\frac{\mu_{\rm crit}}{\mu_i}
\end{equation}
Defining $\delta=\mu_{\rm crit}/\mu_i$, the Dark Matter-Baryon Separability Condition can be written as
\begin{equation}
\frac{\epsilon_\chi}{\epsilon_b}\leq\delta
\end{equation}
For the illustrative values $\mu_i=100$ and $\mu_{\rm crit}=1$ considered previously in our work \cite{trilobdm}, one obtains $\delta=10^{-2}$. A dark matter deficient remnant then requires the formation process to incorporate dark matter at least two orders of magnitude less efficiently than baryons. This formulation allowed the observed deficiency to be mapped onto constraints on dark matter interactions, dissipative components and halo escape. \\

However, the condition we proposed dealt fundamentally with the relative separability of dark matter and baryons, rather than with dark matter deficiency alone. It hence naturally contains a converse branch in which baryons are retained less efficiently than dark matter, but this should not be interpreted as a prediction of the existence of almost-dark galaxies. Since dark matter dominated low luminosity galaxies, including several Local Group satellites, were already known in some instances \cite{ald1simon2007kinematics,ald2strigari2008common,ald3mcconnachie2012observed,ald4walker2009universal,ald5wolf2010accurate,ald6simon2019faintest}. Rather, the point is that the same separability framework developed from dark matter deficient systems also provides a natural language for describing the opposite extreme. To make this statement precise, it is useful to return to the relation for $\mu_f$ before imposing the one sided dark matter deficient inequality and so we define a relative separability parameter
\begin{equation} \label{Q}
\mathcal{Q}\equiv\frac{\epsilon_\chi}{\epsilon_b}=\frac{\mu_f}{\mu_i}
\end{equation}
The quantity $\mathcal{Q}$ contains both possible directions of relative dark matter-baryon evolution. The dark matter deficient systems that motivated our original analysis correspond to $\mathcal{Q}<1$, while $\mathcal{Q}=1$ describes proportional retention of the two components. The converse regime $\mathcal{Q}>1$ corresponds instead to baryons being retained less efficiently than dark matter. The two branches may hence be written as
\begin{equation}
\mathcal{Q}<1
\end{equation}
for preferential dark matter depletion and
\begin{equation}
\mathcal{Q}>1
\end{equation}
for preferential baryonic depletion. This symmetry is mathematical rather than physical, as the mechanisms producing the two outcomes need not be related. Dark matter deficient systems may arise through processes that preferentially remove dark matter or assemble baryons without a corresponding dark matter component. We know that the earth and the sun are dominated by ordinary matter, so the baryons were able to separate from dark matter to make them. Whereas prospective dark matter dominated low luminosity galaxies can instead result from inefficient baryon retention or star formation. In dwarf galaxies, supernova feedback and stellar winds can expel gas from a shallow potential well, while in a cluster environment such as Virgo, ram pressure stripping can further reduce the gaseous baryonic component. The separability parameter therefore classifies the outcome of the evolution without requiring the same physical mechanism to operate on both sides. \\

The two sided character of the framework can be made more transparent by defining a signed separability variable
\begin{equation}
\mathcal{S}_{\chi b}\equiv\ln\mathcal{Q}=\ln\left(\frac{\mu_f}{\mu_i}\right)
\end{equation}
The sign of $\mathcal{S}_{\chi b}$ identifies the direction of the relative change in composition. Dark matter deficient galaxies have $\mathcal{S}_{\chi b}<0$, while systems whose dark matter to baryon ratio is unchanged relative to the progenitor have $\mathcal{S}_{\chi b}=0$, and preferentially baryon depleted systems have $\mathcal{S}_{\chi b}>0$. The magnitude $|\mathcal{S}_{\chi b}|$ measures the degree to which the final dark matter-baryon partition differs from that of the progenitor. Importantly, this mathematical symmetry does not imply equal formation probabilities, comparable evolutionary histories or identical astrophysical mechanisms for the two signs of $\mathcal{S}_{\chi b}$. A generic two sided measure of strong separation may hence be expressed by introducing a chosen separation threshold $\delta_{\rm sep}<1$ and requiring
\begin{equation}
\min\left(\mathcal{Q},\mathcal{Q}^{-1}\right)\leq\delta_{\rm sep}
\end{equation}
which may equivalently be written as
\begin{equation}
\left|\ln\left(\frac{\mu_f}{\mu_i}\right)\right|\geq\ln\left(\frac{1}{\delta_{\rm sep}}\right)
\end{equation}
This form treats the two directions of compositional separation on equal mathematical footing while leaving their physical origin unrestricted. We distinguish $\delta_{\rm sep}$ here from the particular quantity $\delta=\mu_{\rm crit}/\mu_i$ introduced for dark matter deficient systems, since there is no reason in general to impose identical observational thresholds on the two branches. More explicitly, suppose a baryon depleted system is characterized by a final dark matter to baryon ratio satisfying $\mu_f\geq\mu_{\rm high}$. The general relation then gives
\begin{equation}
\frac{\epsilon_\chi}{\epsilon_b}\geq\frac{\mu_{\rm high}}{\mu_i}
\end{equation}
or equivalently
\begin{equation}  
\frac{\epsilon_b}{\epsilon_\chi}\leq\frac{\mu_i}{\mu_{\rm high}}
\end{equation}
This is the mathematical counterpart of the dark matter deficient separability condition, although the astrophysical processes responsible for reaching it can be entirely different. The significance of TTT J1237327+143535 is hence not that it represents the first known dark matter dominated or baryon poor galaxy, nor that its existence was uniquely predicted by the separability framework. Rather, its exceptionally low surface brightness, small stellar mass and location in the Virgo environment make it a particularly interesting new system in which the positive separability branch can be quantified. The broader implication is that dark matter deficient galaxies and baryon depleted galaxies can be placed within a common framework based on relative dark matter-baryon retention, while the mechanisms driving each population are allowed to remain physically distinct.\\

While it is now apparent that our condition can indeed be consistent with the observations of almost-dark galaxies, let us now take this one step further. We shall now use the condition to predict properties of this galaxy which has not yet been observed. The particularly useful missing observable is the line of sight velocity dispersion $\sigma_{\rm los}$ and a measurement of this quantity would determine the enclosed dynamical mass and would therefore allow the almost-dark nature of the system to be distinguished from genuine preferential baryon depletion. Before we do that though, we note that the calculations below have three major caveats. First, we retain the illustrative progenitor value $\mu_i=100$ used in our previous analysis, although the actual progenitor ratio of TTT J1237327+143535 is not known. The choice $\mu_i=100$ will be retained only as an illustrative benchmark, allowing direct comparison with our previous work, rather than as an a priori prediction for TTT J1237327+143535. A future measurement of $\sigma_{\rm los}$ primarily determines the present enclosed dark matter to baryon ratio $\mu_f$, while establishing $\mathcal{Q}>1$ additionally requires comparison with an independently motivated progenitor value of $\mu_i$. Where available, complementary information from the globular cluster population may provide an independent handle on the present halo mass, although it does not by itself determine the progenitor ratio. The relations derived below should therefore be interpreted as a family of conditional and observationally testable consistency relations rather than as predictions based on a unique assumed formation history. Second, we assume that the baryonic mass within the half-light region is dominated by the observed stellar component. The absence of detected HI and H$\alpha$ supports this assumption qualitatively, although the present HI upper limit is not sufficiently restrictive to establish it. Third, we adopt the Virgo distance used in the discovery analysis, which fixes both the inferred stellar mass and physical effective radius. These assumptions make the numerical predictions conditional, but they do not affect the physical soundness of the relations themselves, which can be straightforwardly rescaled when improved observations become available. Also note that although TTT J1237327+143535 provides our immediate worked example because its kinematics remain unmeasured, the resulting relations apply more generally to the wider population of extremely baryon poor and dark matter dominated dwarf galaxies. \\

For a pressure supported galaxy, the dynamical mass near the three dimensional half light radius can be estimated from the projected effective radius as
\begin{equation}
M_{\rm dyn}(<r_{1/2})\simeq\frac{4R_e\sigma_{\rm los}^2}{G}
\end{equation}
The corresponding enclosed dark matter mass is then
\begin{equation}
M_\chi(<r_{1/2})=M_{\rm dyn}(<r_{1/2})-M_b(<r_{1/2})
\end{equation}
and consequently the final enclosed dark matter to baryon ratio becomes
\begin{equation}
\mu_f=\frac{4R_e\sigma_{\rm los}^2}{GM_b(<r_{1/2})}-1
\end{equation}
This identifies what can be obtained directly from a future kinematic measurement as once $\sigma_{\rm los}$ and the enclosed baryonic mass are known, the present value of $\mu_f$ can be inferred without assigning the galaxy to either separability branch beforehand. However, determining whether the system has undergone preferential baryonic depletion requires comparison of this measured $\mu_f$ with the progenitor ratio $\mu_i$, which is not independently known at present. Using the definition $\mathcal{Q}=\mu_f/\mu_i$, the separability parameter inferred from the kinematics is
\begin{equation}
\mathcal{Q}(\sigma_{\rm los},\mu_i)=\frac{1}{\mu_i}\left[\frac{4R_e\sigma_{\rm los}^2}{GM_b(<r_{1/2})}-1\right]
\end{equation}
A velocity dispersion measurement hence determines $\mu_f$ directly, while its interpretation in terms of $\mathcal{Q}$ remains conditional on the progenitor value $\mu_i$. This distinction is important because the kinematic measurement by itself establishes the present dark matter content of the galaxy, whereas separability concerns how that content differs from the composition of its progenitor. To place TTT J1237327+143535 on the positive separability branch, we minimally require $\mathcal{Q}\geq1$
and combining this requirement with the dynamical mass estimator gives
\begin{equation}
\frac{4R_e\sigma_{\rm los}^2}{GM_b(<r_{1/2})}-1\geq\mu_i
\end{equation}
It is easy to check that the corresponding critical line of sight velocity dispersion is
\begin{equation}
\sigma_{\rm crit}(\mu_i)=\left[\frac{GM_b(<r_{1/2})(1+\mu_i)}{4R_e}\right]^{1/2}
\end{equation}
The positive separability branch is hence characterized by $\sigma_{\rm los}\geq\sigma_{\rm crit}(\mu_i)$. This is more appropriately interpreted as a family of conditional dynamical thresholds rather than as a unique velocity dispersion prediction. For a specified progenitor ratio, a dispersion above the corresponding threshold tell us that dark matter has been retained more efficiently relative to baryons than in the progenitor. The observed total stellar mass of TTT J1237327+143535 is approximately \cite{tttle2026serendipitous} 
\begin{equation}
M_\star=2.16\times10^6M_\odot
\end{equation}
Under the assumption that the baryonic mass in the half light region is dominated by the stellar component, we take
\begin{equation}
M_b(<r_{1/2})\simeq\frac{M_\star}{2}=1.08\times10^6M_\odot
\end{equation}
Using this value together with $R_e=0.93\,\mathrm{kpc}$ gives the general threshold
\begin{equation}
\sigma_{\rm crit}(\mu_i)\simeq1.12\sqrt{1+\mu_i}\,\mathrm{km\,s^{-1}}
\end{equation}
This expression is our principal dynamical result for TTT J1237327+143535 under the adopted baryonic and distance assumptions. It leaves the uncertain progenitor ratio very clear and allows any future constraint on $\mu_i$ to be translated immediately into a critical dispersion and the dependence $\sigma_{\rm crit}\propto\sqrt{1+\mu_i}$ also shows explicitly why a unique numerical threshold cannot be quoted without specifying the progenitor composition.

\begin{figure*}[t]
    \centering
    \includegraphics[width=0.8\linewidth]{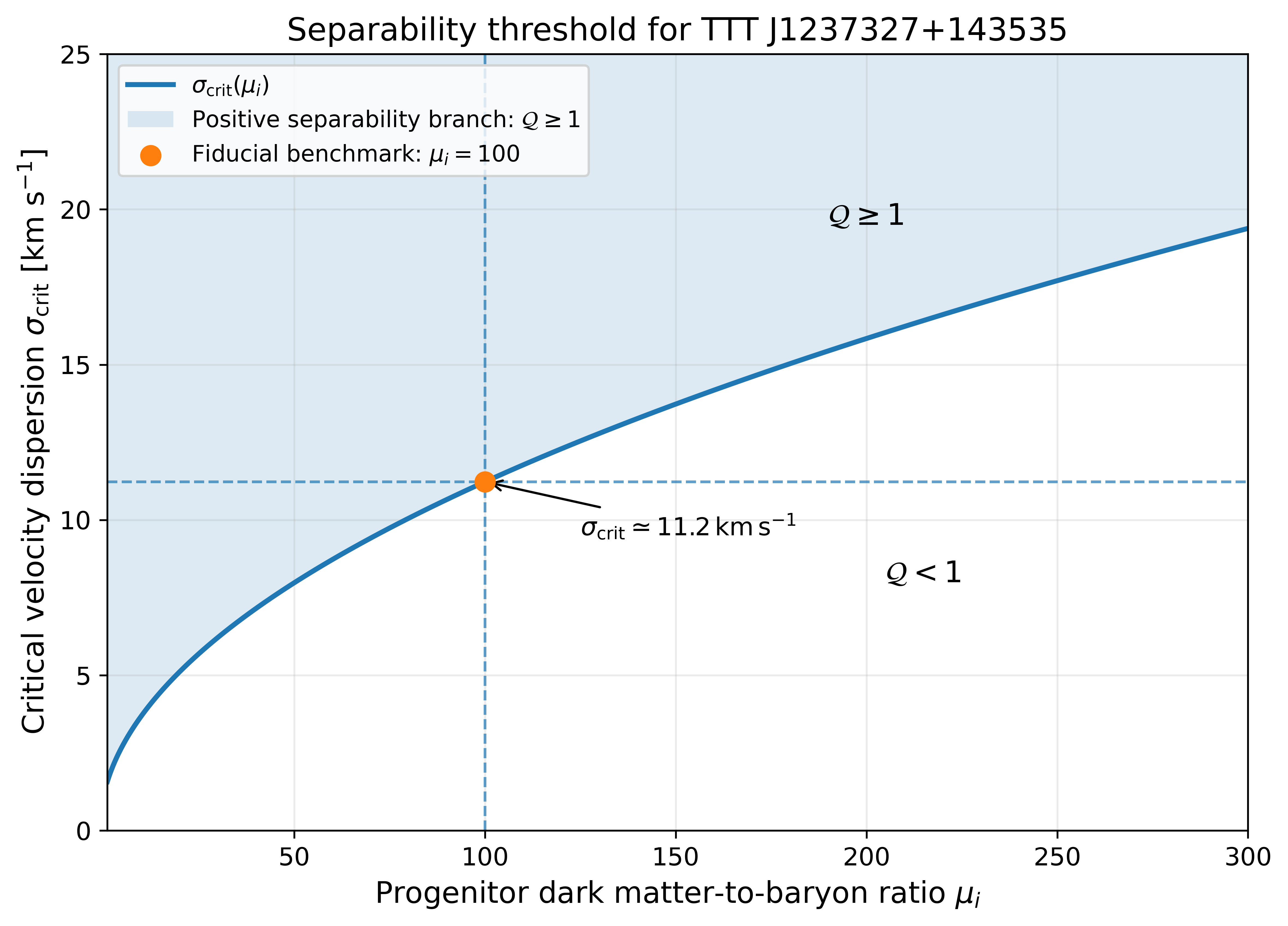}
    \caption{Critical line of sight velocity dispersion for the positive separability branch of TTT J1237327+143535 as a function of the progenitor dark matter to baryon ratio $\mu_i$.}
    \label{vdmsep}
\end{figure*}

It is also useful to write the inferred separability parameter directly in terms of the yet unmeasured velocity dispersion. For the measured $R_e$ and adopted baryonic mass of TTT J1237327+143535, we obtain
\begin{equation}
\mathcal{Q}\simeq\frac{0.801\left(\sigma_{\rm los}/\mathrm{km\,s^{-1}}\right)^2-1}{\mu_i}
\end{equation}
This relation shows that spectroscopy would first determine the numerator, which is simply the inferred present dark matter to baryon ratio $\mu_f$ and the degree of relative dark matter-baryon separation is then obtained only after comparison with $\mu_i$. Thus, future kinematics can map the galaxy onto the separability plane for any independently motivated progenitor ratio rather than merely classify it as dark matter dominated or baryon dominated.For example, retaining $\mu_i=100$ only as an illustrative benchmark, a measured dispersion of $12\,\mathrm{km\,s^{-1}}$ would imply $\mu_f\simeq114$, hence $\mathcal{Q}\simeq1.14$.
Similarly, $\sigma_{\rm los}=15\,\mathrm{km\,s^{-1}}$ would imply $\mu_f\simeq179$, which means $\mathcal{Q}\simeq1.79$. These numbers should be understood only as examples of how a future measurement would be interpreted for a chosen progenitor ratio. The more general result is the continuous relation $\mathcal{Q}(\sigma_{\rm los},\mu_i)$, which separates what is directly measured from what depends on the evolutionary history of the system.
The same reasoning gives a corresponding family of critical enclosed dynamical masses, since
\begin{equation}
M_{\rm dyn}(<r_{1/2})=(1+\mu_f)M_b(<r_{1/2})
\end{equation}
the condition $\mathcal{Q}\geq1$ implies
\begin{equation}
M_{\rm dyn}(<r_{1/2})\geq(1+\mu_i)M_b(<r_{1/2})
\end{equation}
For TTT J1237327+143535 this becomes
\begin{equation}
M_{\rm dyn,crit}(<r_{1/2})\simeq1.08\times10^6(1+\mu_i)M_\odot
\end{equation}
The dynamical mass threshold carries exactly the same dependence on the unknown progenitor ratio as the critical velocity dispersion. Rather than predicting a universal minimum enclosed mass, our separability framework specifies the mass required for the present galaxy to contain a larger dark matter to baryon ratio than its assumed progenitor. For a fiducial value $\mu_i=100$, this reduces to
\begin{equation}
M_{\rm dyn,crit}(<r_{1/2})\simeq1.09\times10^8M_\odot
\end{equation}
More generally, future measurements of $\sigma_{\rm los}$ and the baryonic content will determine $M_{\rm dyn}$ and $\mu_f$ directly, while the comparison of these quantities with plausible progenitor values will determine whether TTT J1237327+143535 genuinely represents the positive branch of dark matter-baryon separability. This is explained more clearly by fig. \ref{vdmsep}, where we see the critical velocity dispersion required for TTT J1237327+143535 to satisfy $\mathcal{Q}\geq1$ as a function of the assumed progenitor ratio $\mu_i$. The shaded region above the curve corresponds to systems whose final dark matter to baryon ratio exceeds that of the progenitor, while the region below the curve has $\mathcal{Q}<1$ and a value of $\sigma_{\rm crit}\simeq11.2\,\mathrm{km\,s^{-1}}$ appears as the special benchmark associated with $\mu_i=100$. This figure makes it clear that future kinematic measurements can be interpreted within a continuous family of separability thresholds once a plausible progenitor ratio is specified. \\

A further consequence of our positive separability branch is concerned with the enclosed baryonic fraction. For this, consider
\begin{equation}
f_b(<r_{1/2})\equiv\frac{M_b(<r_{1/2})}{M_{\rm dyn}(<r_{1/2})}
\end{equation}
Using $M_{\rm dyn}=(1+\mu_f)M_b$, we immediately obtain
\begin{equation}
f_b(<r_{1/2})=\frac{1}{1+\mu_f}
\end{equation}
and since $\mu_f=\mu_i\mathcal{Q}$, this may be written as
\begin{equation}
f_b(<r_{1/2})=\frac{1}{1+\mu_i\mathcal{Q}}
\end{equation}
The baryonic fraction is hence directly linked to the separability parameter and the progenitor dark matter to baryon ratio. In particular, for the positive separability branch $\mathcal{Q}\geq1$, one obtains the general upper bound
\begin{equation}
f_b(<r_{1/2})\leq\frac{1}{1+\mu_i}
\end{equation}
This shows that the stronger the initial dark matter dominance of the progenitor, the smaller the baryonic fraction required for the final system to lie on the positive separability branch. A measured baryonic fraction can hence be compared directly with this threshold once a plausible range of $\mu_i$ is specified. For smaller progenitor ratios the corresponding upper limit is weaker, while larger values of $\mu_i$ demand an even more dark matter dominated final system. Note that this quantity is particularly useful because it probes the partition of the gravitating mass rather than the optical appearance of the galaxy alone. This means that an extremely low surface brightness can arise from a small stellar mass, inefficient star formation or a highly extended stellar distribution, whereas a small value of $f_b$ directly indicates that baryons contribute only a small fraction of the enclosed dynamical mass. A future determination of the baryonic and dynamical masses can hence test whether the almost-dark appearance of TTT J1237327+143535 is accompanied by the degree of baryon depletion required by the positive separability branch. \\

Our framework also provides a corresponding condition on the dynamical mass to light ratio. The reported colour of TTT J1237327+143535 is approximately $(g'-r')_0=0.60$, and the stellar mass to light calibration adopted in the discovery analysis is
\begin{equation}
\log_{10}\left(\frac{M_\star}{L_g}\right)=2.029(g'-r')_0-0.984
\end{equation}
Substituting the observed colour gives
\begin{equation}
\frac{M_\star}{L_g}\simeq1.71\,\frac{M_\odot}{L_{\odot,g}}
\end{equation}
The corresponding total $g$ band luminosity is hence
\begin{equation}
L_g\simeq\frac{2.16\times10^6M_\odot}{1.71M_\odot/L_{\odot,g}}\simeq1.26\times10^6L_{\odot,g}
\end{equation}
Approximately half of this luminosity lies within the projected effective radius, giving
\begin{equation}
L_g(<R_e)\simeq6.3\times10^5L_{\odot,g}
\end{equation}
Under the same assumption used above that approximately half of the stellar mass represents the baryonic mass associated with the half light region, we get
\begin{equation}
\frac{M_b(<r_{1/2})}{L_g(<R_e)}\simeq1.71\,\frac{M_\odot}{L_{\odot,g}}
\end{equation}
This relation is quite useful because the positive separability condition can now be translated directly into a dynamical mass to light threshold. Since $\mathcal{Q}\geq1$ requires $M_{\rm dyn}(<r_{1/2})\geq(1+\mu_i)M_b(<r_{1/2})$, it follows that
\begin{equation}
\frac{M_{\rm dyn}(<r_{1/2})}{L_g(<R_e)}\geq(1+\mu_i)\frac{M_b(<r_{1/2})}{L_g(<R_e)}
\end{equation}
For TTT J1237327+143535, the corresponding general threshold is hence
\begin{equation}
\left(\frac{M_{\rm dyn}}{L_g}\right)_{\rm crit}\simeq1.71(1+\mu_i)\,\frac{M_\odot}{L_{\odot,g}}
\end{equation}
The dynamical mass to light requirement is thus not a unique numerical prediction, but a family of thresholds determined by the assumed progenitor dark matter to baryon ratio. Larger values of $\mu_i$ require a correspondingly larger dynamical mass relative to the observed luminosity for the system to qualify as preferentially baryon depleted. This observable is particularly useful because it combines the extremely small luminosity of TTT J1237327+143535 with an independent measure of its gravitating mass. A large dynamical mass to light ratio would show that the galaxy is not merely optically diffuse, but resides in a gravitational potential substantially more massive than implied by its stellar population alone. When combined with an independently motivated range of $\mu_i$, it can therefore provide another consistency test of whether the system occupies the positive separability branch. \\ 

Finally, another interesting connection can be made when one considers globular clusters. Empirically, the total mass contained in a galaxy's globular cluster system is found to scale approximately linearly with the total dark matter halo mass over a broad range of galaxy masses \cite{ben2026globular}. One can denote this relation by
\begin{equation}
M_{\rm GCS}=\eta_{\rm GC}M_{\rm halo}
\end{equation}
where $\eta_{\rm GC}$ is empirically of order a few times $10^{-5}$. If $\langle M_{\rm GC}\rangle$ denotes the characteristic mass of an individual globular cluster, the expected number of globular clusters can be estimated as
\begin{equation}
\langle N_{\rm GC}\rangle\simeq\frac{M_{\rm GCS}}{\langle M_{\rm GC}\rangle}
\end{equation}
and so,
\begin{equation}
\langle N_{\rm GC}\rangle\simeq\frac{\eta_{\rm GC}M_{\rm halo}}{\langle M_{\rm GC}\rangle}
\end{equation}
\begin{figure*}[t]
    \centering
    \includegraphics[width=0.8\linewidth]{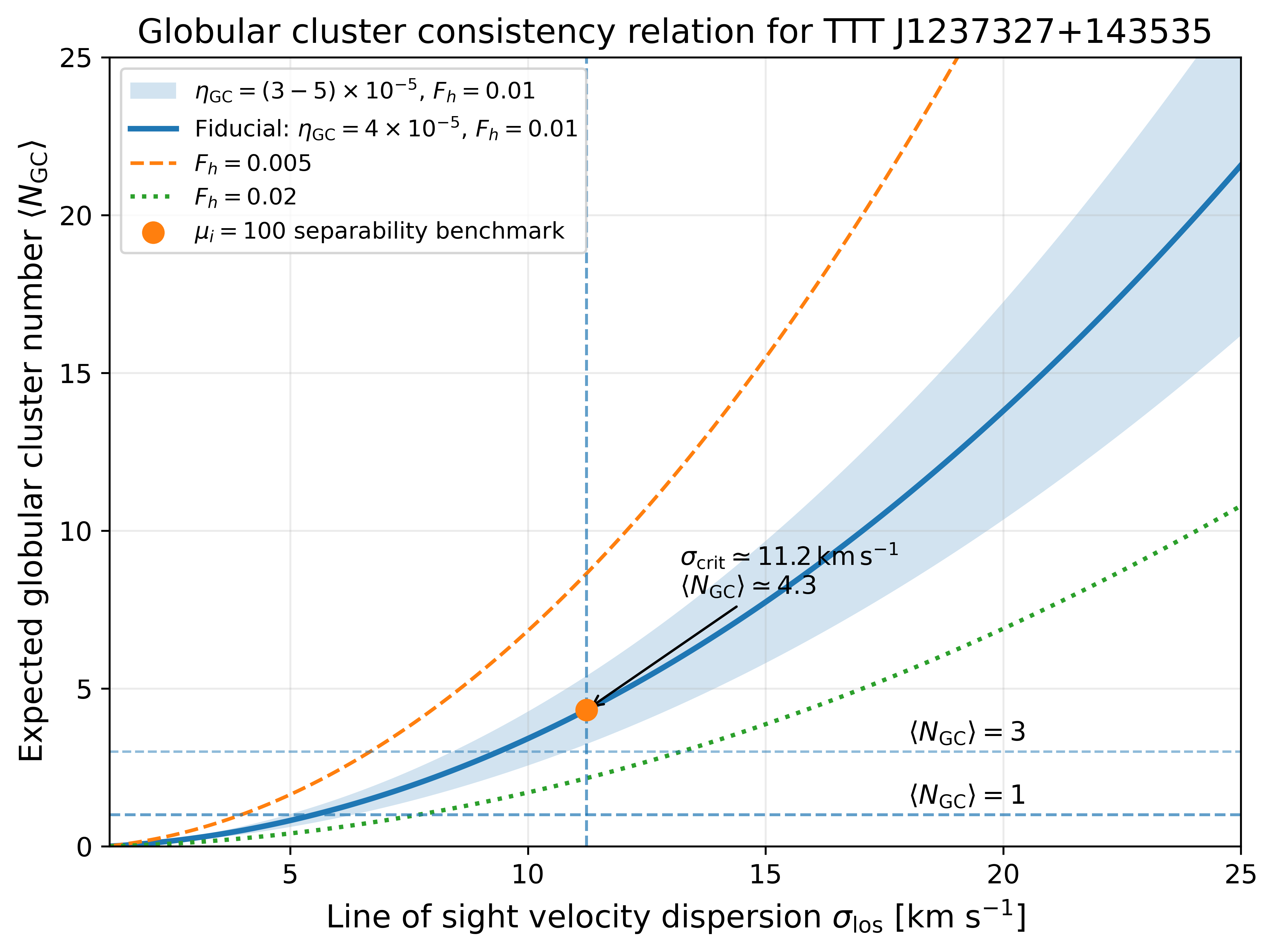}
    \caption{Expected globular cluster population of TTT J1237327+143535 as a function of line of sight velocity dispersion for representative globular cluster scaling and halo structure parameters.}
    \label{gcdmsep}
\end{figure*}
For representative values $\eta_{\rm GC}\simeq(3-5)\times10^{-5}$ and $\langle M_{\rm GC}\rangle\simeq10^5M_\odot$, this corresponds approximately to
\begin{equation}
\langle N_{\rm GC}\rangle\simeq(3-5)\left(\frac{M_{\rm halo}}{10^{10}M_\odot}\right)
\end{equation}
The globular cluster population can hence provide an independent statistical indication of the total halo mass. At sufficiently low halo masses the expected number becomes of order unity or smaller, so the discreteness and intrinsic scatter of the globular cluster population become important.  To connect this result with the separability framework, it is important to distinguish the total halo mass from the dark matter mass enclosed within the half light region, we so define
\begin{equation}
F_h\equiv\frac{M_\chi(<r_{1/2})}{M_{\rm halo}}
\end{equation}
where $F_h$ depends on the halo density profile and its characteristic scale. It follows that
\begin{equation}
M_{\rm halo}=\frac{M_\chi(<r_{1/2})}{F_h}
\end{equation}
Using $M_\chi(<r_{1/2})=\mu_fM_b(<r_{1/2})$ and $\mu_f=\mu_i\mathcal{Q}$ then gives us
\begin{equation}
M_{\rm halo}=\frac{\mu_i\mathcal{Q}M_b(<r_{1/2})}{F_h}
\end{equation}
Substituting this into the globular cluster relation gives
\begin{equation}
\langle N_{\rm GC}\rangle\simeq\frac{\eta_{\rm GC}\mu_i\mathcal{Q}M_b(<r_{1/2})}{F_h\langle M_{\rm GC}\rangle}
\end{equation}
This provides a direct connection between the separability parameter and the expected globular cluster population, although the conversion necessarily depends on the halo structure through $F_h$. The relation does not determine the unknown progenitor ratio $\mu_i$, but it provides an independent consistency check on the present halo mass inferred from the separability analysis.
Now, for the baryonic mass adopted above for TTT J1237327+143535, this relation may be written in a useful normalized form as
\begin{equation}
\langle N_{\rm GC}\rangle\simeq4.3\left(\frac{\eta_{\rm GC}}{4\times10^{-5}}\right)\left(\frac{\mu_i}{100}\right)\mathcal{Q}\left(\frac{0.01}{F_h}\right)\left(\frac{10^5M_\odot}{\langle M_{\rm GC}\rangle}\right)
\end{equation}
The numerical value in this expression should be viewed only as an illustrative normalization, since $F_h$ has not been determined for TTT J1237327+143535. Still, it shows that if the galaxy occupies a sufficiently massive halo, the positive separability interpretation can imply a non-negligible globular cluster population. On the other hand, a robust absence of globular clusters becomes progressively more informative if independent kinematic observations favor a large total halo mass. Interestingly, one can also invert this relation if globular clusters are detected, as from the empirical scaling one has
\begin{equation}
M_{\rm halo}\simeq\frac{\langle N_{\rm GC}\rangle\langle M_{\rm GC}\rangle}{\eta_{\rm GC}}
\end{equation}
which gives an enclosed dark matter to baryon ratio
\begin{equation}
\mu_f\simeq\frac{F_h\langle N_{\rm GC}\rangle\langle M_{\rm GC}\rangle}{\eta_{\rm GC}M_b(<r_{1/2})}
\end{equation}
and consequently
\begin{equation}
\mathcal{Q}\simeq\frac{F_h\langle N_{\rm GC}\rangle\langle M_{\rm GC}\rangle}{\eta_{\rm GC}\mu_iM_b(<r_{1/2})}
\end{equation}
Thus, a measurement of the globular cluster population could provide a second route to the present dark matter content, complementary to the inference from stellar velocity dispersion. The two methods depend on different observables and systematics, making their mutual consistency particularly useful for testing the positive separability interpretation. Combining the globular cluster estimate with the kinematic expression for the enclosed dark matter mass gives the consistency relation
\begin{equation}
\frac{4R_e\sigma_{\rm los}^2}{G}-M_b(<r_{1/2})\simeq\frac{F_h\langle N_{\rm GC}\rangle\langle M_{\rm GC}\rangle}{\eta_{\rm GC}}
\end{equation}
which is equivalently
\begin{equation}
\sigma_{\rm los}^2\simeq\frac{G}{4R_e}\left[M_b(<r_{1/2})+\frac{F_h\langle N_{\rm GC}\rangle\langle M_{\rm GC}\rangle}{\eta_{\rm GC}}\right]
\end{equation}
This relation is particularly useful because it connects two independently accessible observables, the stellar velocity dispersion and the globular cluster population, through the same underlying halo. Agreement between the halo mass inferred from these two routes would strengthen the interpretation of TTT J1237327+143535 as a genuinely dark matter dominated and preferentially baryon depleted system. While on the other hand, a significant discrepancy would instead point toward uncertainties in the halo profile, the low mass globular cluster scaling relation or the evolutionary history of the galaxy. In fig. \ref{gcdmsep} we show how the line of sight velocity dispersion can be mapped onto the expected globular cluster population through the inferred dark matter halo mass. For the fiducial choices $\eta_{\rm GC}=4\times10^{-5}$ and $F_h=0.01$, the expectation rises from $\langle N_{\rm GC}\rangle\simeq0.8$ at $\sigma_{\rm los}=5\,\mathrm{km\,s^{-1}}$ to approximately $4.9$, $7.7$ and $13.8$ at $12$, $15$ and $20\,\mathrm{km\,s^{-1}}$, respectively. The shaded region reflects the empirical range $\eta_{\rm GC}=(3-5)\times10^{-5}$, while the additional curves illustrate the sensitivity to the halo conversion factor $F_h$. The horizontal lines at $\langle N_{\rm GC}\rangle=1$ and $3$ indicate the transition from a regime where a nondetection is statistically unsurprising to one where the absence or presence of several clusters becomes observationally informative. A future determination of both $\sigma_{\rm los}$ and the GC population can hence provide us with two complementary estimates of the underlying halo mass and a nontrivial consistency test of the baryon depleted separability interpretation. Note that we have also have an important caveat here, namely that the outer halo and some of the globular clusters may have been stripped by the tidal force of the Virgo cluster. \\

It is also worth noting that we can get bounds on $\mu_i$ through the tidal properties of the Virgo Cluster, which is particularly relevant for TTT J1237327+143535 given its likely membership in this environment. The tidal survival condition directly constrains the present dark matter to baryon ratio $\mu_f$, while $\mu_i$ enters only after this result is combined with the separability relation $\mathcal{Q}=\mu_f/\mu_i$. Hence, we first derive a lower bound on the present gravitational support and on $\mu_f$, and subsequently connect this result with the separability framework. The basic requirement is that the mean density of the galaxy within a bound radius $r_h$ must exceed the mean density of the Virgo Cluster within the galaxy's cluster centric distance $R$, which we write approximately as
\begin{equation}
\frac{M_g(<r_h)}{r_h^3}>\frac{M_V(<R)}{R^3}
\end{equation}
Here $M_g(<r_h)$ is the total gravitating mass of the galaxy inside its bound extent and $M_V(<R)$ denotes the Virgo mass enclosed within the galaxy's three dimensional cluster-centric distance. This represents the leading tidal survival condition and neglects order unity corrections associated with the detailed orbital configuration and the logarithmic slope of the Virgo mass profile. Only the projected separation from the centre of Virgo is presently known and the observed projected separation from M87 is  then approximately given as
\begin{equation}
R_{\rm p}=0.77\,{\rm Mpc}
\end{equation}
For a spherical and isotropically oriented geometry one has $R_{\rm p}=R\sin\theta$ and taking the statistical geometric deprojection factor gives
\begin{equation}
\left\langle\frac{R}{R_{\rm p}}\right\rangle=\frac{\pi}{2}
\end{equation}
We adopt
\begin{equation}
R\simeq\frac{\pi}{2}R_{\rm p}\simeq1.21\,{\rm Mpc}
\end{equation}
which should be regarded as a representative geometric estimate rather than a direct measurement of the three dimensional position of TTT J1237327+143535. The resulting bounds can be straightforwardly rescaled for another value of $R$ if the true cluster centric distance becomes known. 
\begin{figure*}[t]
    \centering
    \includegraphics[width=0.82\linewidth]{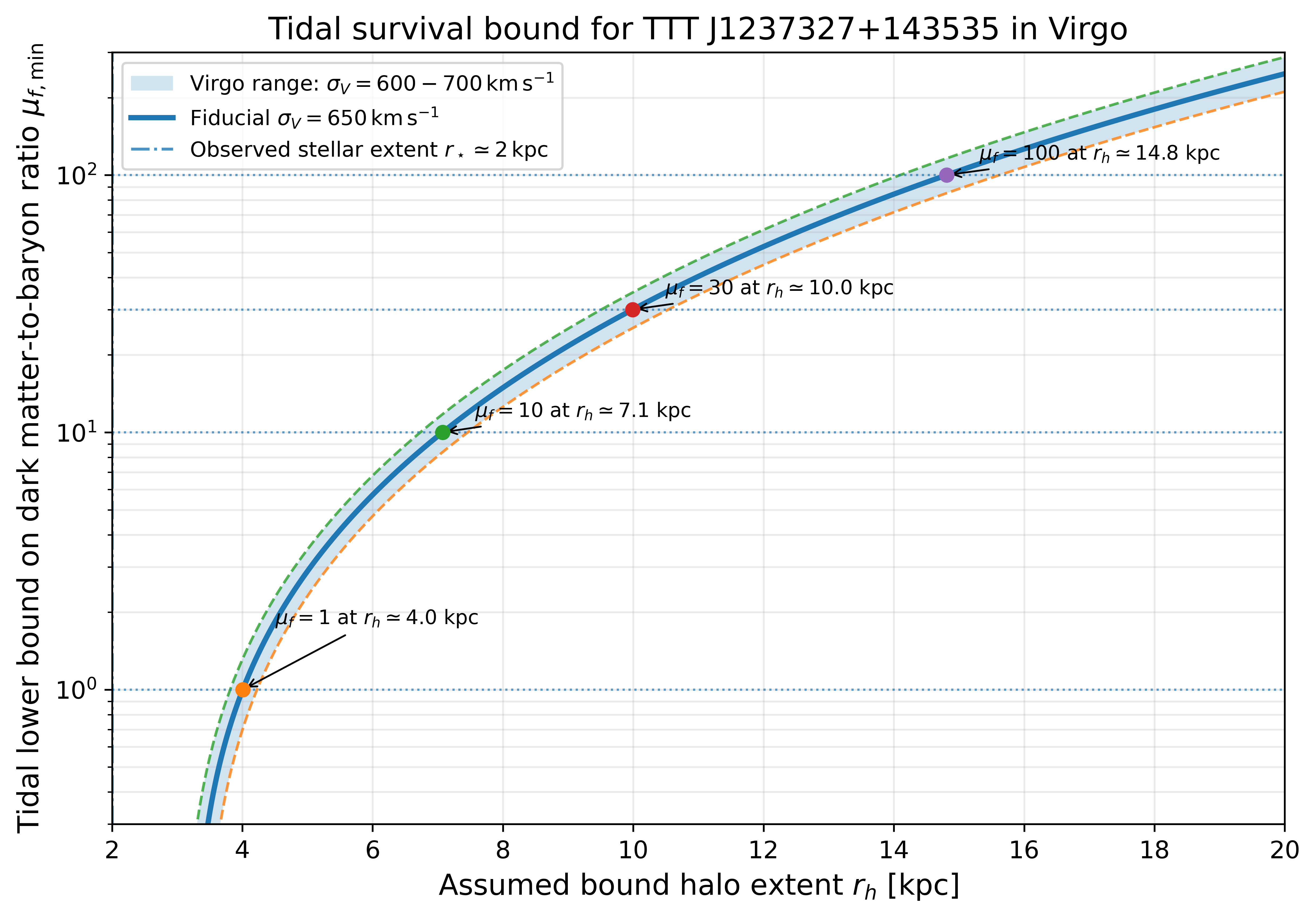}
    \caption{Tidal lower bound on the present dark matter to baryon ratio of TTT J1237327+143535 as a function of its assumed bound halo extent.}
    \label{mutidal}
\end{figure*}
For an approximately dispersion supported gravitating system, we may write
\begin{equation}
\sigma_g^2\simeq\frac{GM_g(<r_h)}{r_h}
\end{equation}
and similarly characterize the Virgo gravitational potential through
\begin{equation}
\sigma_V^2\simeq\frac{GM_V(<R)}{R}
\end{equation}
Substituting these expressions into the tidal survival condition gives
\begin{equation}
\frac{\sigma_g^2}{Gr_h^2}>\frac{\sigma_V^2}{GR^2}
\end{equation}
and so we have
\begin{equation}
\sigma_g>\sigma_V\frac{r_h}{R}
\end{equation}
The Virgo velocity dispersion hence imposes a lower limit on the characteristic gravitational velocity scale of any bound halo extending to $r_h$ and adopting $\sigma_V=600-700\,{\rm km\,s^{-1}}$ and $R\simeq1.21\,{\rm Mpc}$ \cite{td1lebeau2026gas,td2ko2017edge} gives
\begin{equation}
\sigma_{g,\rm min}\simeq(0.496-0.579)\left(\frac{r_h}{\rm kpc}\right)\,{\rm km\,s^{-1}}
\end{equation}
For a representative value $\sigma_V=650\,{\rm km\,s^{-1}}$, this reduces to
\begin{equation}
\sigma_{g,\rm min}\simeq0.537\left(\frac{r_h}{\rm kpc}\right)\,{\rm km\,s^{-1}}
\end{equation}
The observed stellar distribution extends to approximately $r_\star\simeq2\,{\rm kpc}$, which provides a minimum physical scale that the bound system must support. At this radius the tidal condition gives
\begin{equation}
\sigma_{g,\rm min}(2\,{\rm kpc})\simeq0.99-1.16\,{\rm km\,s^{-1}}
\end{equation}
Thus, survival of the currently observed stellar body requires only a characteristic velocity scale of order $1\,{\rm km\,s^{-1}}$. This is a rather weak constraint at the visible stellar radius, although it strengthens linearly if the bound dark halo is shown to extend significantly farther than the observed stars. The same argument can be expressed as a direct lower limit on the enclosed gravitating mass, as from the tidal condition one obtains
\begin{equation}
M_{g,\rm min}(<r_h)=\frac{\sigma_V^2r_h^3}{GR^2}
\end{equation}
Writing the total gravitating mass in terms of the baryonic and dark matter components gives $M_g(<r_h)=M_b(<r_h)\left(1+\mu_f\right)$
and consequently
\begin{equation}
\mu_f>\frac{\sigma_V^2r_h^3}{GR^2M_b(<r_h)}-1
\end{equation}
Since a physical dark matter to baryon ratio cannot be negative, the appropriate tidal lower bound is
\begin{equation}
\mu_{f,\rm min}(r_h)=\max\left[0,\frac{\sigma_V^2r_h^3}{GR^2M_b(<r_h)}-1\right]
\end{equation}
Outside the observed stellar body we may approximately take the enclosed baryonic mass to have saturated at the measured stellar mass $M_\star\simeq2.16\times10^6M_\odot$, subject to the gas caveats discussed previously. For the representative Virgo velocity dispersion $\sigma_V=650\,{\rm km\,s^{-1}}$, the bound can then be written numerically as
\begin{equation}
\mu_{f,\rm min}(r_h)\simeq\max\left[0,0.0311\left(\frac{r_h}{\rm kpc}\right)^3-1\right]
\end{equation}
The cubic dependence on $r_h$ is considerably stronger than the linear dependence of the velocity bound. Consequently, the tidal condition is weak at the observed stellar radius but rapidly becomes restrictive if the galaxy possesses a spatially extended bound halo. The halo radius at which tidal survival first requires a nonzero dark matter contribution follows by setting $\mu_{f,\rm min}=0$ and this gives
\begin{equation}
r_{\rm DM}=\left(\frac{GM_bR^2}{\sigma_V^2}\right)^{1/3}
\end{equation}
which for $\sigma_V=650\,{\rm km\,s^{-1}}$ gives $r_{\rm DM}\simeq3.18\,{\rm kpc}$. This means that the observed stellar body at approximately $2\,{\rm kpc}$ can satisfy the simplified tidal criterion without requiring dark matter. However, if a bound halo is independently shown to extend beyond approximately $3\,{\rm kpc}$, the survival requirement begins to impose a nonzero lower limit on $\mu_f$. \\

For example, the representative $\sigma_V=650\,{\rm km\,s^{-1}}$ case gives approximately $\mu_{f,\rm min}(4\,{\rm kpc})\simeq0.99$, $\mu_{f,\rm min}(7\,{\rm kpc})\simeq9.7$, $\mu_{f,\rm min}(10\,{\rm kpc})\simeq30$ and $\mu_{f,\rm min}(15\,{\rm kpc})\simeq104$.
Allowing the Virgo velocity dispersion to vary from $600$ to $700\,{\rm km\,s^{-1}}$ broadens these values which gives us $\mu_{f,\rm min}\simeq0.70-1.31$ at $4\,{\rm kpc}$, $8.1-11.4$ at $7\,{\rm kpc}$, $25.5-35.1$ at $10\,{\rm kpc}$ and $88-121$ at $15\,{\rm kpc}$. The tidal environment can hence provide a substantial lower bound on the present dark matter content if the spatial extent of the bound halo is established observationally. This result connects naturally back to the separability parameter through \eqref{Q}, from which we can write $\mathcal{Q}\geq\frac{\mu_{f,\rm min}(r_h)}{\mu_i}$ and the tidal argument thus supplies us an environmental lower envelope in separability space for any specified progenitor ratio. In particular, for tidal survival itself to require the positive separability branch $\mathcal{Q}\geq1$, one must have $\mu_{f,\rm min}\geq\mu_i$, which implies
\begin{equation}
r_h\geq\left[\frac{GM_b(1+\mu_i)R^2}{\sigma_V^2}\right]^{1/3}
\end{equation}
For the illustrative value $\mu_i=100$ and $\sigma_V=650\,{\rm km\,s^{-1}}$, this corresponds to
\begin{equation}
r_h\gtrsim14.8\,{\rm kpc}
\end{equation}
Thus, a sufficiently extended bound halo could allow the Virgo tidal field alone to demand a present dark matter to baryon ratio characteristic of the positive separability branch. The more general result is independent of any particular choice of $\mu_i$ and is contained in the lower bound $\mu_{f,\rm min}(r_h)$. \\

In fig. \ref{mutidal} we show the tidal lower bound on $\mu_f$ as the assumed bound halo radius is varied, with the shaded region representing the Virgo velocity dispersion range $\sigma_V=600-700\,{\rm km\,s^{-1}}$. The constraint is negligible close to the observed stellar extent but rises rapidly as $\mu_{f,\rm min}\propto r_h^3$, reaching order unity near $4\,{\rm kpc}$, order ten near $7\,{\rm kpc}$ and order one hundred near $15\,{\rm kpc}$. The representative $\sigma_V=650\,{\rm km\,s^{-1}}$ curve represents the central estimate, while the spread quantifies the sensitivity to the adopted Virgo velocity scale. The figure hence shows that a future determination of the spatial extent of the bound halo can transform tidal survival into a direct lower bound on the present dark matter content.  These tidal constraints are complementary to, rather than replacements for, the dynamical and globular cluster consistency relations derived above. At the presently observed stellar extent the bound is weak, with $\mu_{f,\rm min}=0$ and $\sigma_{g,\rm min}$ of only order $1\,{\rm km\,s^{-1}}$, so it does not exclude the range of $\sigma_{\rm los}$, $M_{\rm dyn}$ or globular cluster populations considered previously. Moreover, $\sigma_g$ in the tidal estimate is a characteristic halo velocity scale and need not coincide exactly with the stellar line of sight dispersion without specifying a halo model. The importance of the tidal argument is instead that it provides an independent environmental lower bound which can become increasingly restrictive if future observations demonstrate that the bound halo extends well beyond the visible stellar component. \\

The discovery of TTT J1237327+143535 thus provides us with a timely opportunity to test the two sided nature of dark matter-baryon separability in a newly identified extreme low surface brightness system. Rather than predicting a unique set of dynamical values, the framework now gives us a family of conditional thresholds parameterized by the progenitor ratio $\mu_i$, with $\sigma_{\rm crit}\simeq1.12\sqrt{1+\mu_i}\,\mathrm{km\,s^{-1}}$, $M_{\rm dyn,crit}(<r_{1/2})\simeq1.08\times10^6(1+\mu_i)M_\odot$, $f_b(<r_{1/2})\leq(1+\mu_i)^{-1}$ and $\left(M_{\rm dyn}/L_g\right)_{\rm crit}\simeq1.71(1+\mu_i)M_\odot/L_{\odot,g}$ under the adopted baryonic and distance assumptions. Future stellar kinematics will determine the present enclosed dark matter to baryon ratio $\mu_f$, while deeper measurements of the gas content, distance and any associated globular cluster population can provide complementary constraints on the total halo mass and test the internal consistency of the separability interpretation. In particular, the globular cluster abundance and, if available, their dynamical evolution can furnish an independent probe of the same halo inferred from stellar kinematics. Agreement between these observables for a plausible progenitor history would support the interpretation of TTT J1237327+143535 as occupying the positive, preferentially baryon depleted branch of the separability framework, while a significant mismatch would instead favor explanations based primarily on inefficient star formation, gas removal or an unusually diffuse stellar distribution. \\

\textbf{Acknowledgements}: The work of OT was supported in part by the Vanderbilt Discovery Doctoral Fellowship. The work of AL is supported in part by the Black Hole Initiative, which is funded by GBMF and JTF.
\bibliography{apssamp}

\bibliographystyle{apsrev4-2}
\end{document}